\documentclass[fleqn, usenatbib, useAMS, letters]{mnras}
\usepackage{times,amsmath,amsfonts,amssymb,xspace}
\usepackage{newtxtext,newtxmath}
\usepackage{graphicx}
\usepackage{amsmath}
\usepackage{paralist,xcolor,xspace}
\usepackage{times}
\usepackage{comment}
\usepackage{orcidlink}
\usepackage[super]{nth}
\usepackage[T1]{fontenc}
\usepackage{bold-extra}
\usepackage{bm}
\usepackage{graphicx}
\usepackage{hyperref}

\newcommand{\flamingo}{\textsc{Flamingo}\xspace}
\newcommand{\VR}{\textsc{VELOCIraptor}\xspace}
\newcommand{\msun}{{\rm M}_{\odot}}
\newcommand{\lcdm}{$\Lambda$CDM\xspace}

\DeclareRobustCommand{\VAN}[3]{#2}

\title[The anistropic distribution of local clusters]{On the anisotropic distribution of clusters in the local Universe}

\author[M. Schaller]{
Matthieu Schaller\textsuperscript{\orcidlink{0000-0002-2395-4902}}\thanks{E-mail: \url{mschaller@lorentz.leidenuniv.nl}}\\
Lorentz Institute for Theoretical Physics, Leiden University, PO Box 9506, NL-2300 RA Leiden, The Netherlands\\
Leiden Observatory, Leiden University, PO Box 9513, NL-2300 RA Leiden, The Netherlands
}

\begin{document}  

\pagerange{\pageref{firstpage}--\pageref{lastpage}} 

\pubyear{2023}
\date{Accepted 2023 December 19. Received 2023 December 10; in original form 2023 October 09}

\maketitle

\label{firstpage}


\begin{abstract}
In his 2021 lecture to the Canadian Association of Physicists Congress,
P.J.E. \citeauthor{Peebles2021} pointed out that the brightest extra-galactic
radio sources tend to be aligned with the plane of the de Vaucouleur Local
Supercluster up to redshifts of $z=0.02$ ($d_{\rm MW}\approx 85~\rm{Mpc}$). He
then asked whether such an alignment of clusters is anomalous in the standard
\lcdm framework. In this letter, we employ an alternative, absolute orientation
agnostic, measure of the anisotropy based on the inertia tensor axis ratio of
these brightest sources and use a large cosmological simulation from the
\flamingo suite to measure how common such an alignment of structures is. We
find that only 3.5\% of randomly selected regions display an anisotropy of their
clusters more extreme than the one found in the local Universe's radio
data. This sets the region around the Milky Way as a $1.85\sigma$
outlier. Varying the selection parameters of the objects in the catalogue, we
find that the clusters in the local Universe are never more than $2\sigma$ away
from the simulations' prediction for the same selection. We thus conclude that
the reported anisotropy, whilst note-worthy, is not in tension with the \lcdm
paradigm.
\end{abstract}




\section{Introduction}
\label{sec:intro}

Over the last two decades, the standard cosmological model, the \lcdm model, has
received much scrutiny and passed a multitude of tests \citep[see
  e.g.][]{Dodelson2020, Lahav2022}. This vast program, designed to stress test
the model and understand its limitations, will continue in this decade with
exceedingly demanding precision tests, generally grouped under the Stage IV
cosmology probe label. These are especially designed to help shed some light on
the nature of both dark matter and dark energy as well as to explore some of the
tensions currently emerging between orthogonal probes \citep[see
  e.g.][]{Abdalla2022}.

One of the core tenets of our cosmology paradigm, which has arguably received
comparatively less attention over the last decades, is the assumption of
homogeneity and isotropy of our Universe. This Copernican assumption, coupled to
Einstein's general relativity equations, allows us to fully describe the
evolution and expansion of the Universe. Testing it is challenging, but,
recently, tentative signs of tension have been reported in the literature
\citep[e.g.][]{Migkas2020, Migkas2021, Secrest2022, Kumar2023, Watkins2023}.
Dropping or altering this foundational assumption would, of course, have
dramatic consequences on the interpretation and analysis of many other datasets
and would likely require a complete redesign of our standard cosmological model.

The distribution of galaxies, clusters, or other bright sources in the local
Universe is among the different tentative observational signs of anisotropy. By
using the local volume as a laboratory, one can obtain clean and complete
samples, which can be more challenging on larger scales. Additionally, these
local probes also have the advantage of having a long history and unlike
larger-scale measurements, they directly test whether \emph{our} place in the
Universe is special. \\

\noindent One such anomaly was reported by \cite{Peebles2021}. In his review of
the \lcdm paradigm and some of its problems, he observed that the distribution
of radio sources in the local Universe ($0.01<z<0.02$) was not uniform on the
sky. More specifically, following the exact definitions introduced later in
\cite{Peebles2022}, he noted that the 32 brightest radio sources associated with
galaxies in the all-sky catalogue of \cite{Velzen2012} and in the redshift range
quoted above are found to be strongly aligned with the plane of the de
Vaucouleurs Local Supercluster. As the volume surveyed contains $\sim10^4$
galaxies, this is potentially a clear sign of alignment of massive galaxies on
relatively large scales ($\approx 170~\rm{Mpc}$). A similar signal was already
reported by \cite{Shaver1989}, also in the analysis of radio data and
observations in other bands, in particular X-ray, show the same trend
\citep{Peebles2022}. In his follow-up study, \citet{Peebles2022_paper} altered
slightly his selection but reached the same conclusions.

The discussion of this apparent anomaly by \cite{Peebles2022} then continues by
asking two important questions. He, first, wondered why the clusters of galaxies
appear to be near a plane at $z<0.02$ and, secondly, whether this arrangement is
unlikely within a \lcdm framework, which he seems to suggest. He finally mention
that this latter point could easily be verified by looking at large-scale
structure $N$-body simulations.

In this letter, we directly answer this question by performing exactly the test
proposed above using a catalogue of haloes extracted from a recent large \lcdm
simulation. These cover a volume large enough to contain thousands of distinct
regions allowing us to estimate the likelihood of observing such an anisotropic
distribution of bright sources in a simple frequentist way.


\section{Local cluster anisotropy}
\label{sec:data}

We start by reproducing the analysis of radio galaxies performed by
\cite{Peebles2022}. To be concrete, we pick one of the selections used in the
literature, which we detail here for completeness. We show below that altering
slightly the parameters of the selection does not change the main conclusion.

\begin{figure}
\includegraphics[width=0.9\columnwidth]{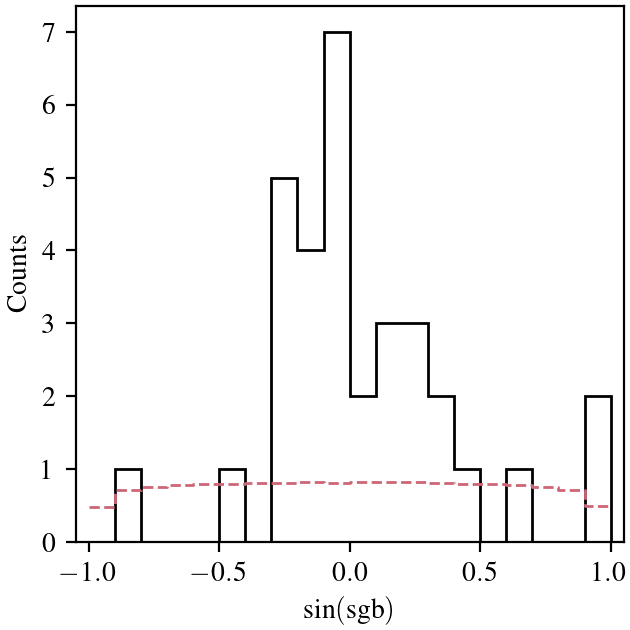}
\caption{The 32 brightest extra-galactic radio sources from the van Velzen
  catalogue \citet{Velzen2012} of radio sources in the redshift range
  $0.01<z<0.02$ binned by the sine of their super-galactic latitude (sgb). The
  dashed red histogram at the bottom corresponds to a uniform distribution
  (arbitrarily normalised) of objects with source, in the zone of avoidance
  (ZoA), i.e. with Galactic $|b|<10^\circ$ removed. Our selection matches the
  left panel of figure~1 of \citet{Peebles2022}. As previously reported, there
  is a clear anisotropy in the distribution of clusters in the local ($z<0.02$)
  Universe. The anisotropy seen in the radio catalogue is not due to the
  presence of the ZoA.}
\label{fig:Peebles_angles}
\end{figure}

We use the \cite{Velzen2012} radio catalogue and extract the extra-galactic
sources within the redshift range $0.01<z<0.02$. For a fiducial cosmology with a
Hubble constant $H_0 = 70~{\rm km/s/Mpc}$, this corresponds to a spherical shell
with inner- and outer-radii of $d_{\rm MW}\approx45$ and $\approx85~\rm{Mpc}$
respectively. We then select the $32$ brightest objects in this shell. The
distribution of these sources in regular bins of the sine of their
super-galactic latitude (sgb) is shown as the black histogram on
Fig.~\ref{fig:Peebles_angles}. This should be compared to an isotropic
distribution. However, the data suffers from potential anisotropies caused by
the difficulties arising from observing through the plane of the Milky
Way. Following \cite{Peebles2022}, we thus show an isotropic distribution of
sources from which a 10-degree ``zone of avoidance'' (ZoA) in Galactic latitude
has been subtracted as the dashed red line on Fig.~\ref{fig:Peebles_angles}. As
reported by \cite{Peebles2022}, the $32$ brightest radio sources are
preferentially aligned with the super-galactic plane ($\rm{sgb}=0$). See also
the left panel of his Fig.~1 which displays the same data, confirming that we
constructed the same sample of objects. The observed anisotropy is not caused by
the presence of the Galactic ZoA. \\

Having established the presence of an anisotropy in the local distribution of
clusters, we now turn to the construction of a quantitative metric of the
anisotropy.

To characterize the anisotropy in absolute terms, \textit{i.e.} without
reference to a particular direction, we borrow the tools used for the analysis
of planes of satellites around the Milky-Way \citep[e.g.][]{Pawlowski2015,
  Sawala2023} and base our analysis around the reduced inertia tensor generated
by the distribution of sources:
\begin{equation}
I_{ij} = \sum_{n=1}^N \mathbf{x}_{n,i} \mathbf{x}_{n,j},
\label{eq:inertia}
\end{equation}
where $\mathbf{x}_n$ are the coordinates of the $n$-th cluster with respect to
the observer after projection onto a unit sphere. We choose to use the reduced
tensor so as to not be affected by distant outliers; this definition is also
close to the idea of angle on the sky used in the original \cite{Peebles2022}
analysis. We label the square roots of the tensor's eigenvalues (\textit{i.e.}
the three axis of the ellipsoid of inertia) as $a$, $b$ and $c$ (in decreasing
size) and define the anisotropy as the ratio $c/a$. A perfectly isotropic
distribution of points would display a $c/a$ ratio of one whilst a perfectly
planar distribution would have $c/a=0$.

Applying this procedure to the 32 brightest sources selected by
\cite{Peebles2022} from the \cite{Velzen2012} radio catalogue in the distance
range $0.01<z<0.02$, we find an anisotropy $c/a_{\rm data} = 0.464$. For
completeness, we also computed the intermediate-to-major axis ratio $b/a_{\rm
  data} = 0.713$. This indicates that the distribution of radio sources is an
oblate, flattened distribution.


\section{Simulations}
\label{sec:sims}

The simulation used for our analysis is taken from the Virgo Consortium's
\flamingo\footnote{\url{https://flamingo.strw.leidenuniv.nl/}} suite
\citep{Schaye2023, Kugel2023}. These simulations, performed using the
\textsc{Swift} cosmology code \citep{Schaller2023}, assume a standard flat
\lcdm cosmology with Gaussian initial conditions and parameters $h=0.681,
\Omega_{\rm m} = 0.306, \Omega_\Lambda = 0.694, n_s= 0.967, \sigma_8 = 0.807$,
which corresponds to best-fit \lcdm model to the Dark Energy Survey year
three data combined with external constraints (the model labelled ``$3\times$2pt +
All Ext.'' of \citet{DES}). More specifically, we make use of the main
``dark matter only'' simulated volume (\texttt{L2p8\_m9\_DMO}) which evolves
$5040^3$ cold dark matter particles in a volume of $2.8^3~{\rm Gpc}^3$. This
combination leads to a particle mass of $6.72\times10^9 {\rm M}_\odot$,
sufficient to fully resolve haloes with a mass of $10^{12} {\rm M}_\odot$,
\textit{i.e.} more an order of magnitude lower than the haloes in which the
sources discussed in the previous section are expected to reside.  Note that for
the purpose of identifying the alignment of large structures on scales of 10s of
Mega-parsec, a simulation evolving only dark matter is sufficient as baryonic
and astrophysics processes won't affect structures on these scales
\citep{Schaye2023}. We use the redshift zero output of the simulation and
identify haloes and their centres in the snapshot using the phase-space
structure finder \VR \citep{Elahi2019}.


\section{Results and analysis}

We now perform an analysis of the simulations to obtain the frequency of systems
as anisotropic as the ones observed in the local Universe.

\subsection{Direct comparison}
\label{ssec:direct}

To mimic the source catalogues introduced above, we place $N=5000$ observers
randomly in the simulation volume\footnote{Using close-packing there are
$\approx5800$ independent spheres in our simulation volume. We use $5000$
observers to avoid repetitions.}. Note that we do not explicitly select
observers to lie on a specific matter sheet, as is the case of the MW which
resides on the sheet of the local Supercluster. We are thus asking what the
anisotropies would look like for a truly random observer rather than for the
ones embedded in a local structure. By design, this choice increases the chance
that our local environment is anomalous and is hence a good basis for a
null-test.

For each observer, we construct an all-sky shell and include all objects at a
distance $45~\rm{Mpc}<r<85~\rm{Mpc}$ of the observer. To mock the zone of
avoidance created by the plane of the Milky Way disk, we draw a random vector
for each of these observers and define the observer's galactic disk as the plane
perpendicular to this vector. The key underlying assumption here being that the
orientation of the plane of the MW disk is in no way related to any large-scale
plane of clusters or filaments. We then eliminate from our source catalogues all
the objects within $10^\circ$ of this plane. Finally, we select the centres of
the 32 largest haloes in each of the regions as a proxy for the selection of the
32 brightest radio sources.  From their location, we construct the reduced
inertia tensor (eq.~\ref{eq:inertia}) of each of these mock catalogues and
compute their anisotropies $c/a$.

\begin{figure}
\includegraphics[width=0.9\columnwidth]{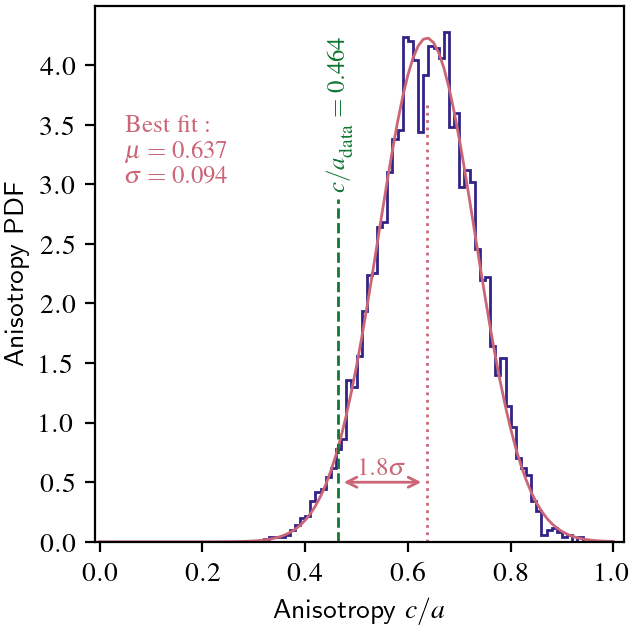}
\caption{Distribution of the anisotropy of the 32 most-massive clusters in 5000
  randomly selected regions of size $45~{\rm Mpc}<r<85~{\rm Mpc}$ extracted from
  the FLAMINGO simulation volume (blue histogram). The solid red line is a
  Gaussian fit to the distribution whose mean is indicated by the vertical
  dotted line and parameters are given in the figure. The green vertical dashed
  line indicates the anisotropy of the sample of 32 radio sources of
  Fig.~\ref{fig:Peebles_angles}. The observed anisotropy in the data is a
  $1.85\sigma$ outlier towards more anisotropic systems.}
\label{fig:anisotropy_distribution_radio}
\end{figure}

The distribution of anisotropies is shown in
Fig.~\ref{fig:anisotropy_distribution_radio}. The distribution of anisotropies
is exceptionally well fit by a Gaussian with a mean $c/a$ of $0.637$ and a
standard deviation of $0.094$
\footnote{A K-S test between the bins and the fit returns a p-value of
$<10^{-6}$.}.  The vertical dashed line indicates the anisotropy measured for
the local catalogue of radio sources. $96.5\%$ of all random mock observers see
a distribution on the sky of their brightest 32 objects within
$45~\rm{Mpc}<r<85~\rm{Mpc}$ that is more isotropic than the one we obtained in
the local Universe. At face value, this puts the region around the MW as a
$1.85\sigma$ outlier towards more anisotropic distributions.

For completeness, we also measured the intermediate-to-major axis ratio ($b/a$)
using the same inertia tensor for all our virtual observers. We find a
distribution of $b/a$ axis ratios with mean $0.812$ and standard deviation
$0.087$, indicating that, similarly to the local Universe data, the virtual
observers see oblate distribution of sources.

As \cite{Pawlowski2015} pointed out, valuable information can be lost when
considering only the reduced inertia tensor and not the full tensor. We have
repeated the analysis of both the data and simulation using the full inertia
tensor and have obtained virtually indistinguishable results ($c/a_{\rm
  data}=0.467$ and a Gaussian distribution in the simulation characterized by a
mean $c/a=0.631$ and a standard deviation of $0.095$). We interpret this
minimal difference as being the natural consequence of the narrow range of
distances considered in the original selection of radio sources by
\citet{Peebles2022}.

\subsection{Properties of the objects probed}

As we started from a catalogue of bright radio sources which we compare to a
simulation without baryonic physics, it is interesting to check whether the
haloes we used are sensible hosts of radio galaxies.  For each of the $5000$
random observers, we recorded the least and most massive objects they
encountered in the radial shells they used for the anisotropy calculations. For
all observers, we find that the largest of the 32 objects included in the
selection lies in the range $M_{200}=1.0\times10^{14}~\msun -
5\times10^{15}~\msun$. Similarly, the \emph{smallest} object they encountered
lies in the range $M_{200}=3.5\times10^{13}~\msun - 1.5\times10^{14}~\msun$. At
the resolution of our simulation, these are all extremely well resolved
haloes. They are also the hosts of large groups of galaxies or even large
clusters. We expect such haloes to all host large central galaxies which will be
radio bright \citep[see e.g.][]{Mandelbaum2009} and would clearly be part of the
selection made by \cite{Velzen2012} had they been observing these regions of the
simulation.

As an additional cross-check, the range of halo masses quoted above is also in
agreement with the estimates of masses made for the objects in the local
Universe \citep[see e.g.][]{Stopyra2021}.

\subsection{Varying the parameters of the problem}

The analysis presented above relies on a specific choice for the radial range
used and for the number of objects to retain when computing the angular
distribution or anisotropy. To strengthen our conclusions, we repeat, here, the
analysis by varying these aforementioned parameters.

\begin{figure}
\includegraphics[width=\columnwidth]{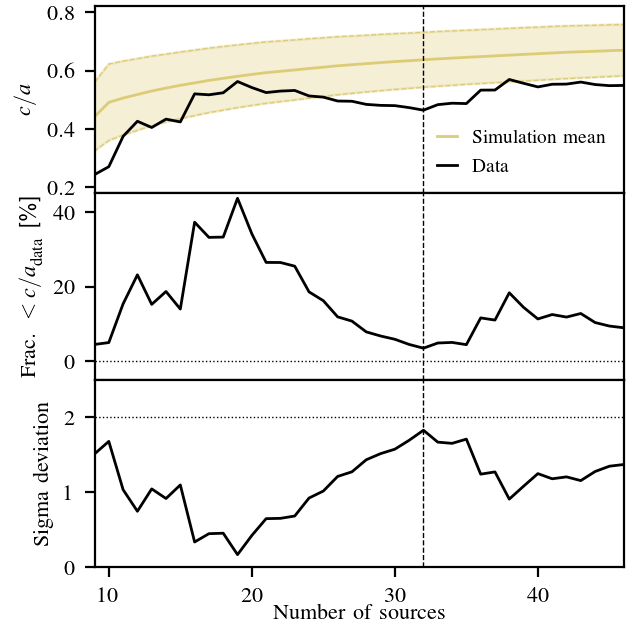}
\caption{A repeat of the analysis for varying number of sources in the
  catalogue. \textit{(top)} The anisotropy of the top $N$ brightest sources in
  the van Velzen catalogue in the radial range $45~\rm{Mpc}<r<85~\rm{Mpc}$
  (black) and the mean and standard deviation of the distribution obtained from
  5000 random observers in the simulation (yellow). \textit{(middle)} The
  fraction of regions of the same size in the simulation that have a flatter
  distribution of their top $N$ most massive clusters. \textit{(bottom)} The
  deviation from the simulation's mean of the observed $c/a$ in units of the
  standard deviation of the distribution.  On all panels, the dashed line
  indicates the number of sources chosen by \citet{Peebles2022} and corresponds
  to the results shown in Fig.~\ref{fig:anisotropy_distribution_radio}. For no
  choice of the number of sources does the observed local anisotropy deviates at
  the $2\sigma$ level or more from the simulation-inferred \lcdm expectation.}
\label{fig:look_elsewhere_N}
\end{figure}

We start by keeping the radial range fixed but changing the number of objects to
consider. Instead of choosing the $32$ brightest sources in the radio catalogue,
we let this number float. For each selection, we compute the corresponding
inertia tensor and reduced anisotropy. This latter value is reported as a
function of the number of sources on the top panel of
Fig.~\ref{fig:look_elsewhere_N}. We then return to the simulation and for each
of the $5000$ random observers, we compute the anisotropy based on the top $N$
largest haloes in the radial range $45~{\rm Mpc}<r<85~{\rm Mpc}$ after removal
of a mock zone of avoidance (see above). The fraction of observers with an
anisotropy of their clusters more pronounced than the data is reported in the
middle panel of Fig.~\ref{fig:look_elsewhere_N}. Finally, for each choice of
$N$, we compute the mean and standard deviation of $c/a$ and report the distance
from the mean of the observed one in units of the standard deviation on the
bottom panel of Fig.~\ref{fig:look_elsewhere_N}.

We find that for all values of $N$, the deviation from the \lcdm
expectation, given by the simulation, of the observed anisotropy of local
clusters never reaches a $2\sigma$ threshold.

\begin{figure}
\includegraphics[width=\columnwidth]{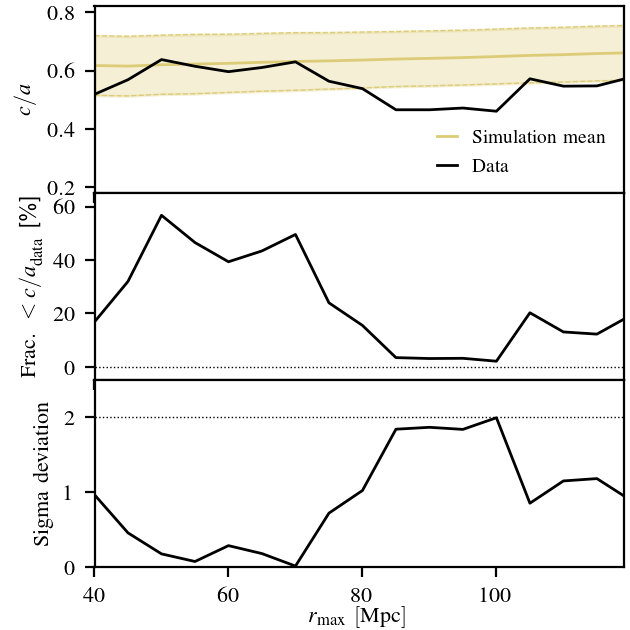}
\caption{Same as Fig.~\ref{fig:look_elsewhere_N} but varying the radial range
  used to select the clusters. \textit{(top)} The anisotropy of the top 32
  brightest sources in the van Velzen catalogue in the radial range between a
  minimum of $30~\rm{Mpc}$ and a maximum given on the x-axis (black) and the
  mean and standard deviation of the same measurement across 5000 random
  observers in the simulation (yellow). \textit{(middle)} The fraction of
  regions of the same size in the simulation that have a flatter distribution of
  their top 32 most massive clusters. \textit{(bottom)} The deviation from the
  simulation's mean of the observed $c/a$ in units of the standard deviation of
  the distribution.  For no radial range choice does the observed local
  anisotropy deviates at the $2\sigma$ level or more from the
  simulation-inferred \lcdm expectation.}
\label{fig:look_elsewhere_r}
\end{figure}

The other variable in the problem is the choice of radial range away from the
Milky Way where the radio sources (or clusters) are selected. We explore this
now by varying the thickness of the radial shell from which the clusters are
selected. We keep the inner radius fixed and vary the maximum one. In order to
always have enough sources, we reduce the inner radius to $r_{\rm min} = 30~{\rm
  Mpc}$. For each shell thickness, we select the 32 brightest sources in the
catalogue or the 32 most massive haloes in the simulation. The result of the
experiment as a function of the shell outer-radius $r_{\rm max}$ is displayed in
Fig.~\ref{fig:look_elsewhere_r}. As was the case when varying the number of
sources, we find that the anisotropy found in the data never reaches a $2\sigma$
deviation from the simulation's mean.

It is interesting to notice that the data is found to be most anisotropic
when compared to the simulation's mean when taking a shell reaching up to
$r_{\rm max} = 80 - 100~{\rm Mpc}$. This is the distance at which the Perseus
supercluster is located.


\section{Conclusion}
\label{sec:conclusion}

In this letter, we directly answer the question asked by \citet{Peebles2021}
where he wondered whether the observed distribution of massive clusters in the
very local Universe ($z<0.02$) is in tension with the standard cosmological
model.

To achieve this, we used one of the large \lcdm $N$-body simulations of the
\flamingo suite \citep{Schaye2023} to study the isotropy of the distribution of
clusters in spheres of radius $\approx 70~\rm{Mpc}$. More specifically, we made
mimicked the selection of bright radio sources, done by \cite{Peebles2022}, that
revealed the anisotropic distribution of clusters in the Local Universe. By
placing $5000$ random observers in the simulation, each constructing the same
mock survey, we were able to construct the anisotropy of clusters in shells of
$45~\rm{Mpc} < r < 85~\rm{Mpc}$ for a \lcdm cosmology with Gaussian initial
conditions. We then used the reduced inertia tensor, constructed from the
clusters' centres, and its eigenvalues as a measure of the isotropy of the
clusters in each region and found that the anisotropy identified in the Local
Universe data (Fig.~\ref{fig:Peebles_angles}) is a $1.85\sigma$ outlier
(Fig.~\ref{fig:anisotropy_distribution_radio}) from the \lcdm-based
prediction. In other words, $96.5\%$ of the regions in the simulation display an
arrangement of their $32$ most massive clusters in the aforementioned radial
shell that is more isotropic than what the radio data of \cite{Velzen2012}
reveals.

We then varied the parameters of the selection by changing the number of objects
selected (Fig.~\ref{fig:look_elsewhere_N}) and the radial range
(Fig.~\ref{fig:look_elsewhere_r}). In both cases, we found that the
observational data never reaches a $2\sigma$ threshold compared to the
simulation-inferred \lcdm-prediction. \\

We thus conclude that the anisotropy in the distribution of bright sources
revealed by \cite{Peebles2021, Peebles2022, Peebles2022_paper} in the all-sky
radio data of \cite{Velzen2012} and \cite{Shaver1989} (as well as in survey data
using other wavelengths) is not a problem for the \lcdm model.


\section*{Acknowledgments}
We thank Carlos Frenk, Jim Peebles, and Till Sawala for useful comments on an
early draft of this manuscript as well as the \flamingo collaboration
for making their simulation data available.\\
This work used the DiRAC@Durham facility managed by the Institute for
Computational Cosmology on behalf of the STFC DiRAC HPC Facility
(\url{www.dirac.ac.uk}). The equipment was funded by BEIS capital funding via
STFC capital grants ST/K00042X/1, ST/P002293/1, ST/R002371/1 and ST/S002502/1,
Durham University and STFC operations grant ST/R000832/1. DiRAC is part of the
National e-Infrastructure.

\section*{Data availability}
The data from the \flamingo suite will be made public once practically feasible
given the challenging size of the datasets. In the mean time, the data can be
accessed upon requests to the authors.

\bibliographystyle{mnras}
\DeclareRobustCommand{\VAN}[3]{#3}
\bibliography{./bibliography.bib}

\label{lastpage}

\end{document}